\documentstyle[12pt,a4]{article}
\input{epsf.tex}
\begin{document}
\begin{center}
{\large {\bf Intense two-step cascades and $\gamma$-decay scheme of 
the $^{118}$Sn compound nucleus}}
\end{center}
\begin{center}{\bf JAROSLAV. HONZ\'ATKO$^a$,
VALERY. A. KHITROV$^b$, CVETAN PANTELEEV$^b$, ANATOLY. M. SUKHOVOJ$^b$, IVO.
TOMANDL$^a$}\\
{\sl $^a$ Nuclear Physics Institute, CZ-25068 \v{R}e\v{z} near Prague, Czech Republic}\\
{\sl $^b$Frank Laboratory of Neutron Physics, Joint Institute for Nuclear
Research \\ 141980 Dubna, Russia}
\end{center}

\begin{center}
UDC 539.172.4
\end{center}
\begin{center}
\end{center}

Transition energies and intensities of 454 two-step cascades following
thermal neutron capture in $^{117}Sn$ have been measured. These data allowed
us to make more precise and considerably extend the earlier obtained decay
scheme of the $^{118}Sn$ nucleus.

Autocorrelation analysis of the excitation spectrum of intermediate levels
of the most intense cascades allowed us to determine the most probable period
of their regularity. Besides, the analysis showed impossibility quantitatively
 to reproduce
 the total cascade intensity with accounting only for the
excitations of the fermion type. This resulted into a  conclusion that the excitations
of a probable vibrational type in $^{118}Sn$ also considerably influence the
cascade $\gamma$-decay process  practically up to the neutron binding energy
$B_n$.

\section{Introduction}
Both understanding of the processes occurring in a nucleus and precise
calculation of important for practice parameters of these processes (for
example, neutron cross-sections) requires their detailed experimental study
in the wide excitation energy interval and in the large mass range of nuclei.

In these frameworks, the two-step $\gamma$-cascades following thermal neutron
capture in target nucleus $^{117}Sn$ have been studied. The energies of
$\gamma$-transitions and intensities of most strong 454 cascades exciting 
303 levels in the compound nucleus were determined in the energy interval 
practically up to $B_n$. 

\section{Experiment}
The experiment was performed at LWR-15 reactor in \v{R}e\v{z} [1].
The target, enriched in $^{117}Sn$ to more than 91\%,  was used in the
experiment.
The contribution of other $Sn$ isotopes
in the background of $\gamma -\gamma$ coincidences was very small.

The sum coincidence spectrum of the two-step cascades measured in the
experiment is presented in Fig.~1. Using this spectrum, intensity
distributions for the cascades terminating at 8 low-lying levels with
$E_f< 2.8$ MeV of $^{118}Sn$ were built.
Each two-step cascade in such spectrum is presented by a pair of peaks
with equal [2] areas and widths. Energy resolutions in each of these
distributions is varying  from 1.9 keV at their ends to 3 keV in centre.
Such a resolution at the achieved statistics of coincidences, low and
practically constant at any energy background in these spectra,
which is  provided the
detection threshold $\simeq 0.6\times 10^{-4}$ events per decay for the
cascades to the ground state and $2 \div 5\times 10^{-4}$ events per decay for the
cascades at the higher-lying levels.

The spectrum in Fig.~1 is very specific as compared with those for other 
nuclei, studied by us earlier. It has single-escape and double-escape peaks
 corresponding not only to one cascade transition but also to both 
 transitions simultaneously. Thus, cascades to the ground and first excited
  states of $^{118}Sn$ create in the sum coincidence spectrum four 
well-expressed additional peaks shifted by 511 keV.

\section{Decay scheme}
The confidence of the data  on peculiarities of the nuclear excited states
determining intensities of the observed two-step cascades is provided,
first of all, by the reliability and completeness of the obtained scheme
of $\gamma$-decay. Two-step cascades following neutron capture unambiguously 
set energies of the initial
and final levels but do not contain direct information on the energy of
their intermediate level. However, if cascade is intense enough
($i_{\gamma\gamma}\geq 10^{-4}$ events per decay) then it is usually
observed as a pair of the resolved peaks. In this case one can determine
the quanta ordering in cascade.

The method to construct a decay scheme using obvious thesis about the
constancy of the energy $E_1=B_n-E_i$ of the primary transition in the
cascades with the different total energy $E_1+E_2=B_n-E_f=const$ was described
for the first time in [3]. The method uses the multi-dimensional Gauss
distribution in the framework of the maximum likelihood method in order
to select probable $\gamma$-transitions with equal energy in different
spectra. As it was shown in [4], the algorithm gives reliable results
even at the mean error $\simeq 1$ keV in determination of the transition
energies for some hundreds of cascades placed into the decay scheme.
Corresponding decay scheme for $^{118}Sn$ is given in Table~1.

Relative intensities of most intense cascades for all 8 spectra obtained in
the experiment were transformed into absolute values by normalization
to the absolute intensities of their primary transitions [5] (Table 2) and
branching ratios obtained in traditional manner from the mass of coincidences
accumulated in the experiment. The involving of the maximal number of the most
intense cascades in normalization procedure decreases its uncertainty due
to correlations between them. Some notion of the values of statistical and
systematic errors is given by a comparison of the sum of intensities of
all cascades proceeding via the same intermediate level with the experimental
intensity of their primary transitions. At the absence of
systematic errors in determination of cascade intensities, the
ratio $R = \sum i_{\gamma\gamma}/i_1$  should decrease when $E_1$ decreases.
Exceeding of $R$ over the unity gives some total uncertainties for both
intensities. Besides, owing to sufficiently different backgrounds in the
spectra of cascade transitions and single $\gamma$-transitions, contribution
of systematic error of $i_1$ in this uncertainty is larger than that of
$i_{\gamma\gamma}$ at least for the smallest $E_1$.

The total absolute intensities $I_{\gamma\gamma}=\sum{i_{\gamma\gamma}}$
of cascades with a fixed sum energy (including those unresolved
experimentally) obtained in this way at the detection threshold
of quanta $E_{\gamma}>520$ keV are listed in Table~3. Here energy
interval of the cascade final levels $E_f$ is limited by conditions of the
experiment --  at the higher energies $E_f$ the ratio peak/background
in the sum coincidence spectrum decreases. This does not allow one to
get reliable information on the function $i_{\gamma\gamma}=f(E_1)$
for $E_f>2-3$ MeV  without the Compton-suppressed spectrometer.

The total intensities  $I_{\gamma\gamma}$ are suitable for revealing the
serious discrepancies between the experiment and model notions of level density
and radiative strength functions. Results of calculation of
$I_{\gamma\gamma}^{mod}$ with the use of level density model [6], model of
radiative strength functions [7] for E1- transitions and strength function
k(M1)=const
 are listed in Table 3. Considerable exceeding of the experimental
cascade intensities over the calculated values means that the level density
really excited after the thermal neutron capture is less than predictions of
model [6] and, probably, that the radiative strength functions of the primary
transitions have more strong energy dependence than that predicted by
model [7].

\subsection{Comparison with a known decay scheme}

The decay scheme of $^{118}Sn$ includes 454 two-step cascades proceeding
via 303 intermediate levels with the excitation energy up to
$\simeq (B_n-1)$ MeV; 110 levels from them are depopulated by two or more
secondary transitions. Hence, reality of these levels is confirmed with a
high confidence. (The mean uncertainty in determination of the energy
of the cascade quantum is 0.48 keV).  Quanta ordering for 191 cascades
cannot be determined within algorithm [3] because corresponding intermediate
levels are depopulated only by one secondary transition.
In general case, many of these cascades can have  primary transition with
lower energy than energy of the secondary transition.
Quanta ordering in these 191 cascades was determined using the data
ENDS-file [8] on the estimated decay scheme of the nucleus under study.

It should be noted that the levels 2120, 2577 and 2725 keV listed in this
file were not observed in our experiment neither in the sum coincidence
spectrum nor as the intermediate levels of other cascades. Therefore, using
data [8] one should take into account that the estimated decay scheme can
contain wrong levels with mistaken spin values at this and, moreover, at
higher energy.

Neutron capture cross section in a target is mainly determined by the only
neutron resonance with the known parameters. Therefore, one can unambiguously
assign spin and parity  $J^{\pi}=1^+$ to the compound state of $^{118}Sn$.
So, levels of this nucleus excited by the cascades (see Table~1) have, most
probably, both positive and negative parity and spin values in the interval
0-2. The absence in the sum coincidence spectrum of the peak corresponding
to registration of the two-step cascades to level $E_f=2280$ keV,
$J^{\pi}=4^+$ [8] shows that, as in nuclei studied earlier, the total
intensity of cascades with one purely quadrupole transition in $^{118}Sn$
is less than the detection threshold of the experiment.

\subsection{Fluctuations of the cascade intensities and problem of
completeness of the decay scheme}

According to the conventional notions, partial widths of primary transitions
following decay of neutron resonance of non-magic nucleus are the random
values whose fluctuations with respect to the average obey the Porter-Thomas
distribution [9]. Unfortunately, this notion was not verified for both
primary $\gamma$-transitions to the levels with the energy about several MeV
and in the region of the smallest partial widths. Hence, non-Gaussian
character of the distribution of some amplitudes of the 
primary $\gamma$-transitions
of a given multipolarity cannot be excluded. This can be due to, for example,
contribution of large components of the wave function of the levels connected
by corresponding transition in the matrix element of the primary
$\gamma$-transition. In this case, experimental distribution of small partial
widths will have less probability than it follows from model [9] owing to
violation of the condition of application of the central limit theorem of
the mathematical statistics.

The necessity  of this question in practical solution arises at the analysis
of the experimental data on the cascade intensities. Very low and rather even
background in intensity distributions of the cascades to the lowest levels
of the studied nucleus provides better sensitivity for the experiment on the
study of coincidences relative to the traditional experiment [5] with a
single detector. This problem for $^{118}Sn$ is more important than, for
example, for deformed nuclei due to bigger number of intermediate levels, 
which are depopulated by the only $\gamma$-transition with the intensity
exceeding registration threshold.

The presence of 191 cascades of this kind and, correspondingly, the necessity
to determine energies of their intermediate levels (without using method [3])
requires one to solve the dilemma:

(a)~either there are many levels at relatively low excitation energies in
$^{118}Sn$ (for instance, $E_i< 3-4$ MeV), which do not manifest themselves
in other reactions, and, as a consequence, level density in this nucleus is
noticeably higher than even the predictions of model [6], at least,
in this energy interval;

(b)~or the number of primary transitions whose widths are 5-10 times 
smaller
than the mean value is considerably less than expectation according to [9]
for the same excitation energy interval.

Considerably less threshold value of the function
approximating the distribution for $E_m > 5.5$ MeV in comparison with this
parameter for lower energies admits only two these possible interpretations.
This conclusion follows from Fig.~2, which represents the comparison between the
cumulative sums of the experimental intensities of cascades proceeding via
the same intermediate level (for the energy bin $\Delta E_i=0.5$ MeV) and
 corresponding approximation (as it was done for the first time in [10]).
Experimental data in Fig.~2 were obtained under condition that the 
low-intensity cascades were placed into the decay scheme in such a manner
that their primary transitions have the lower energy. This assumption 
was used only when it was impossible to use for this aim the method [3] or 
data on excited levels from [8].

\section{Cascades with the low-energy primary $\gamma$-transitions}
Fig.~3 represents a part of the experimental intensity distribution
of the cascades to the ground state of $^{118}Sn$. Decay scheme of this
nucleus up to $E_{ex}\geq 2.5$ MeV was very reliably established in different
experiments. Based on all available data one can state to a precision of 
about 100\% that this spectrum contains in  the interval 520 to 2500 
keV only 4 peaks corresponding to registration of the low-energy secondary transitions [8] of cascades. 
The rest of the spectrum  should be related with the
registration of the cascades with the energies  of primary transitions
$E_1 < 2.5$ MeV.
The majority of the cascades with the high-energy primary transitions are
usually registered in resolved intense peaks, and a number of low-intensity
cascades with low-energy primary transitions form pseudo-continuous distribution in
the spectrum.

But observation of the resolved peak related with registration of possible
low-energy primary transition of corresponding cascade permits one
to get some information on structure of intermediate level of cascade.
 Intensity of individual cascade
proceeding via a given intermediate level:
\begin{equation}
i_{\gamma\gamma}=i_{1}\times {i_2/\sum i_2}
\end{equation}
is always less than the intensity of its primary transition  $i_1$ because
the branching ratio $b_r=i_2/\sum i_2$ cannot be greater than the unity.
Unfortunately, background under the peaks in the sum coincidence spectrum
increases when the total energy of cascade $E_1+E_2$ decreases. This makes
impossible the extraction of the main part of the secondary transition
intensities at the decay of any high-energy enough intermediate level and,
as a consequence, to determine $i_1$ as a sum of all values $i_{\gamma\gamma}$
for a given intermediate level. But for some probable primary transitions one
can select from [5] close on the energy $\gamma$-quantum (which in many cases
was not placed into the decay scheme suggested by authors or placed in such a
manner that it was not observed in corresponding distribution of the cascade
intensities). Radiative strength functions determined for such cases can be
compared with the predictions of model [7]. 
Analysis [11], made for the
experimental intensity distributions  $I_{\gamma\gamma}$ of two-step cascades
to the ground and first excited states of $^{118}Sn$ with accounting for 
the known total radiative width  $\Gamma_{\lambda}=80$ meV, shows that the 
radiative strength function up to the energy $E_1 \simeq 3.5$ MeV does not 
exceed the value predicted by model [7]. Therefore, this model can be used 
for estimation of probability of low-energy primary transitions whose partial
widths considerably exceed the mean value.

In this experiment have been observed, at least, 15 cascades with the most
intense, low-energy primary transitions ($E_1 < 2.5$ MeV) whose partial
widths are, at least, 20 times larger than the prediction of [7]. According
to model [6], approximately 26000 levels are expected to be excited by the
primary $E1$ and $M1$ $\gamma$-transitions in the excitation energy region
from 6.5 to 8.3 MeV. The probability to observe the only random exceeding
of partial width over the average by a factor of 20 and more for the
considered $\gamma$-transitions in the framework of the Porter-Thomas
distribution is equal to $10^{-5}$.

Hence, from the above statement that some of the observed in [5]
$\gamma$-transitions are really primary follows that some part of these
transitions have anomalous large widths whose appearance cannot be random.
Therefore, some of the levels with the excitation 
energy $\geq 6$ MeV must considerably
differ in the structure of the wave functions from the neighbouring levels.
I. e., the effects like those observed at the low excitation energy manifest
themselves and at higher excitations.

The presence of very intense low-energy primary $\gamma$-transitions was established earlier [12] in the $^{126,127,129,131}Te$ compound nuclei. They are interpreted as direct capture of neutron at the sub-shells $3P_{3/2}$ and $3P_{1/2}$. Unfortunately, clarifying of possible structure of the considered
levels  of $^{118}Sn$ is impossible without further experimental and theoretical investigations.

\section{Possible regularity of the excitation spectrum of the
intermediate levels of the  most intense cascades}

According to the modern theoretical notions, the wave function structure of
any excited states is determined by co-existence and interaction between the
fermion (quasi-particles) and boson (phonons) excitations. With the excitation
energy a nucleus transits from practically mono-component excitations of the
mentioned types to the mixed (quasi-particles $\otimes$ phonons) states with
rather different [13] degree of their fragmentation. This process should be
investigated in details but there is no adequate experimental methods to
study the structure of the wave functions, at least, up to the excitation
energy not less than 5 MeV.

Nevertheless, some information on the probable dominant components of wave
functions of heavy deformed nuclei can be obtained even in this case.
The authors of [14] suggested to search for the regularity in the
excitation spectra of the intermediate levels of the most intense cascades by
means of auto-correlation analysis of the smoothed distributions of
the sum cascade intensities from Table~1. Intensities were smoothed
by means of the Gaussian function:
$F(E)=\sum i_{\gamma\gamma} \times exp(-0.5(\Delta E/\sigma)^2)$.
The distribution of this type smoothed with the parameter $\sigma=25$ keV
is given in Fig.~4, and the values of the auto-correlation function
\begin{equation}
A(T)=\sum_{E}F(E)\times F(E+T)\times F(E+2T)
\end{equation}
for different selection thresholds of intense cascades are shown in Fig.~5.
As it was shown in [15], such an analysis cannot give unique value of the
equidistant period $T$ even for the simulated spectra (for example, for 25
``bands" consisting from 4 levels with slightly distorted equidistant period
and intensities of cascades)
and provide estimation of the confidence level of the observed effect. In
principle, both problems can be solved in the experiments on the study of
the two-step cascades in different resonances of the same nucleus.
But some grounds to state that the regularity really exists can be obtained
from a comparison of the most probable equidistant periods in different
nuclei. The set of the probable equidistant periods obtained so far (Fig.~6)
allows an assumption that the $T$ value is approximately proportional to the
number of boson pairs of the unfilled nuclear shells. This allows one
to consider the effect at the level of working hypothesis, though the
probability of random existence  of the regular spectrum of intermediate
levels of the most intense cascades in each, arbitrary nucleus cannot
be equal to zero.

The regularity in the excitation spectra testifies to the harmonic
nuclear vibrations.
 Thus, one can assume that the structure of the
intermediate levels of the studied cascades contains considerable components
of the rather weakly fragmented states like multi-quasi-particle excitations
$\otimes$ phonon or several phonons. This provides logical explanation of
serious decrease in the observed level density as compared with the
predictions of the non-interacting Fermi-gas model: nuclear excitation
energy concentrates on phonos but quasi-particle states in the energy
interval up to $\simeq 2$ MeV are excited weakly or very weakly due
to insufficiency of energy for breaking of paired nucleons.
 This is quite probable but, in principle, not the only explanation.

\section{Cascade population of levels in $^{118}Sn$.}

Obvious deviations of level density and radiative strength functions
from predictions of from the simplest models like [6,7] and
probable regularity in the excitation spectra of intermediate levels of the
most intense cascades show that the notions of the cascade $\gamma$-decay
process need development and some corrections. This necessity follows from
a comparison between the experimental and model calculated values of the
population of the high-lying levels of the nucleus under study. It is seen
from eq.~(1) that at the presence of reliable information on the intensities
of individual cascades and known [5] intensities of their primary and
secondary transitions one can to determine $\sum i_2$, which equals the
total population of a given level by the direct primary transitions and any
possible cascades. It is worthwhile to subtract the primary transition
intensity from this value in order to decrease its fluctuation. The
populations obtained in this manner for the levels with positive parity are 
compared with the model calculated values in Fig.~7. It should be noted that 
difference between the calculated populations of levels with the same spin 
but different parities is insignificant.
 It is seen that the correspondence between the experiment
and calculation for the levels above $\approx 3$ MeV cannot be achieved.
On the one side, this result confirms the conclusion made above on the presence
of the excitation energy interval where occurs principle change in properties
of the excited states of not only deformed but also spherical nuclei and,
on the other side, it points at the necessity to develop more precise
nuclear models for the all interval of the excitation energies
$E_m \simeq B_n$.

\section{Conclusion}
For the first time, the large enough and reliable scheme of the excited
states and modes of their decay for the $^{118}Sn$ compound nucleus were
obtained in the experiment with the excitation energy  more than 3 MeV.

Analysis of this information shows, in particular, that an improvement of
precision in description of properties of spherical nucleus requires more
detailed accounting for co-existence and interaction of quasi-particle and 
phonon components of wave functions of the levels excited after the neutron 
capture.
\\\\

This work was supported by GACR under contract No. 202/03/0891 and by RFBR
Grant No. 99-02-17863.

\begin{center}{References} \end{center}
\begin{flushleft}
\begin{tabular}{r@{ }p{5.65in}}
$[1]$ & J. Honz\'atko  et al., Nucl. Instr. and Meth. {\bf A376} (1996) 434\\
$[2]$ & A.M. Sukhovoj and V.A. Khitrov, Sov. J.: Prib. Tekhn. Eksp.
{\bf 5} (1984) 27\\
$[3]$ & Yu.P. Popov, A.M. Sukhovoj, V.A. Khitrov and Yu.S. Yazvitsky,
Izv. AN SSSR, Ser. Fiz. {\bf 48} (1984) 1830\\
$[4]$ & S.T. Boneva, E.V. Vasilieva and  A.M. Sukhovoj, Izv. RAN., Ser. Fiz.,
{\bf 51(11)}  (1989) 2023\\
$[5]$ & Yu.E. Loginov, L.M. Smotritskij, P.A. Sushkov,
Vopr. Atom. Nauki i Tech., ser. Fiz. Yad. Reac.  (in Russian),
{\bf 1/2} (2001) 72\\
$[6]$ & W. Dilg, W. Schantl, H. Vonach and M. Uhl, Nucl. Phys. {\bf A217}
(1973) 269\\
$[7]$ & S.G. Kadmenskij, V.P. Markushev and W.I. Furman,
Sov. J. Nucl. Phys. {\bf 37} (1983) 165\\
$ [8]$ &  K. Kitao, Nucl. Data Sheets, {\bf 75(1)} (1995) 99\\
$    $ & http://www.nndc.bnl.gov/nndc/nsr/nsrframe.html\\
$[9]$  & C.F. Porter and R.G. Thomas, Phys. Rev. 1956. {\bf
104}, (1956) 483\\
$ [10]$ &  A.M.Sukhovoj and V.A. Khitrov, Physics of Atomic Nuclei,
{\bf 62(1)} (1999) 19\\
$ [11]$ & E.V. Vasilieva, A.M.Sukhovoj and V.A. Khitrov,
Physics of Atomic Nuclei, {\bf 64(2)} (2001) 153\\
$ [12]$ & I.Tomandl, V.Bondarenko, D.Bucurescu, J.Honz\'atko, T.von Egidy,
H.-F.Wirth, G.Graw, R. Hertenberger, A.Metz, Y.Eiserman,
Proc. 10 International Symposium on Capture gamma-ray spectroscopy and related topics, Santa Fe, USA,
Ed. by S.Wender, AIP, (1999) 200\\
$[13]$ & L.A. Malov and V.G. Soloviev, Yad. Fiz., {\bf 26(4)} (1977) 729\\
$[14]$ & A.M. Sukhovoj and V.A. Khitrov,
Bulletin of the Russian Academy of Sciences, Physics {\bf 61(11)} (1997)
1611\\
$[15]$ & E.V. Vasilieva et al., Bulletin of the Russian Academy of Sciences,
Physics {\bf 57} (1993) 1582\\
\end{tabular}
\end{flushleft}
\begin{center}
\newpage
{\sl Table ~2.\\
 The energies $E_{\gamma}$  and absolute intensities (\% per decay)
$i_1$ of the most intense transitions used for normalization of the cascade
intensities in $^{118}Sn$. $\sum i_{\gamma\gamma}$ is the observed
intensity of the cascades with corresponding primary transition.
}\end{center}
\begin{center}
\begin{tabular}{|l|l|l|}  \hline
$E_1$, keV& $i_1$ [5]    &
$\sum i_{\gamma\gamma}$\\ \hline
 8096.37(17)  & 0.23(3)  & 0.37(2)\\
 7268.84(13)  & 0.40(3)  & 0.44(1)\\
 6997.7(4)    & 0.37(8)  & 0.31(2)\\
 6648.52(7)   & 1.00(6)  & 0.75(3)\\
 6422.14(11)  & 1.04(9)  & 0.88(4)\\
 6110.0(2)    & 0.99(2)  & 0.80(3)\\
 6055.66(18)  & 0.52(2)  & 0.55(3)\\
 5444.9(3)    & 0.55(7)  & 0.35(1)\\
 5298.1(5)    & 0.33(5)  & 0.28(1)\\
 5217.51(10)  & 0.95(6)  & 0.60(3)\\
 5208.7(2)    & 1.1(1)   & 1.09(4)\\
 4781.55(16)  & 0.79(6)  & 0.28(2)\\
 4765.99(15)  & 0.80(5)  & 0.51(3)\\
 4556.31(14)  & 1.33(6)  & 0.39(4)\\
\hline
Sum    & 9.85(21) &7.6(1) \\ \hline \end{tabular} \end{center}

\begin{center}
{\sl Table ~3.
 Energies $E_1+E_2$ of cascades and their absolute intensities
$I_{\gamma\gamma}$ (\% per decay).
$E_f$ is the energy of the cascade final level}
\end{center}
\begin{center}
\begin{tabular}{|l|r|r|l|}  \hline
$E_1+E_2$, keV &  $E_f$, keV &
$I_{\gamma\gamma}^{exp}$ & $I_{\gamma\gamma}^{mod}$ 
\\ \hline
 9326.30  &0&    16.0(34)&  6.7 \\
 8096.63  &1230& 15.3(11)&  7.2\\ 
 7568.00  &1758&   2.4(7)&  1.1\\
 7283.42  &2042&   3.3(16)& 2.4\\
 7269.39  &2057&   2.8(9)&  0.8\\
 7000.36  &2325+2328& 5.6(9)& 2.8\\
 6923.08  &2403&   2.8(2)&  1.5 \\
 6829.42  &2497&   [2]   &  0.5\\
 6648.95  &2677&   [1.5] & 1.0 \\
 6588.29  &2738&   [4]   & 1.7 \\\hline
sum       &    & 55.8(43)& 25.7 \\
\hline
\end{tabular} \end{center}
Note:
intensity of the cascades to the levels $E_f=$ 2497, 2677 and 2738 keV
was estimated from a comparison of their peak areas in the sum coincidence
spectrum with the peak area corresponding to the cascades to the levels
2325 and 2328 keV.

\begin{figure}
\begin{center}
\leavevmode
\epsfxsize=12.5cm
\epsfbox{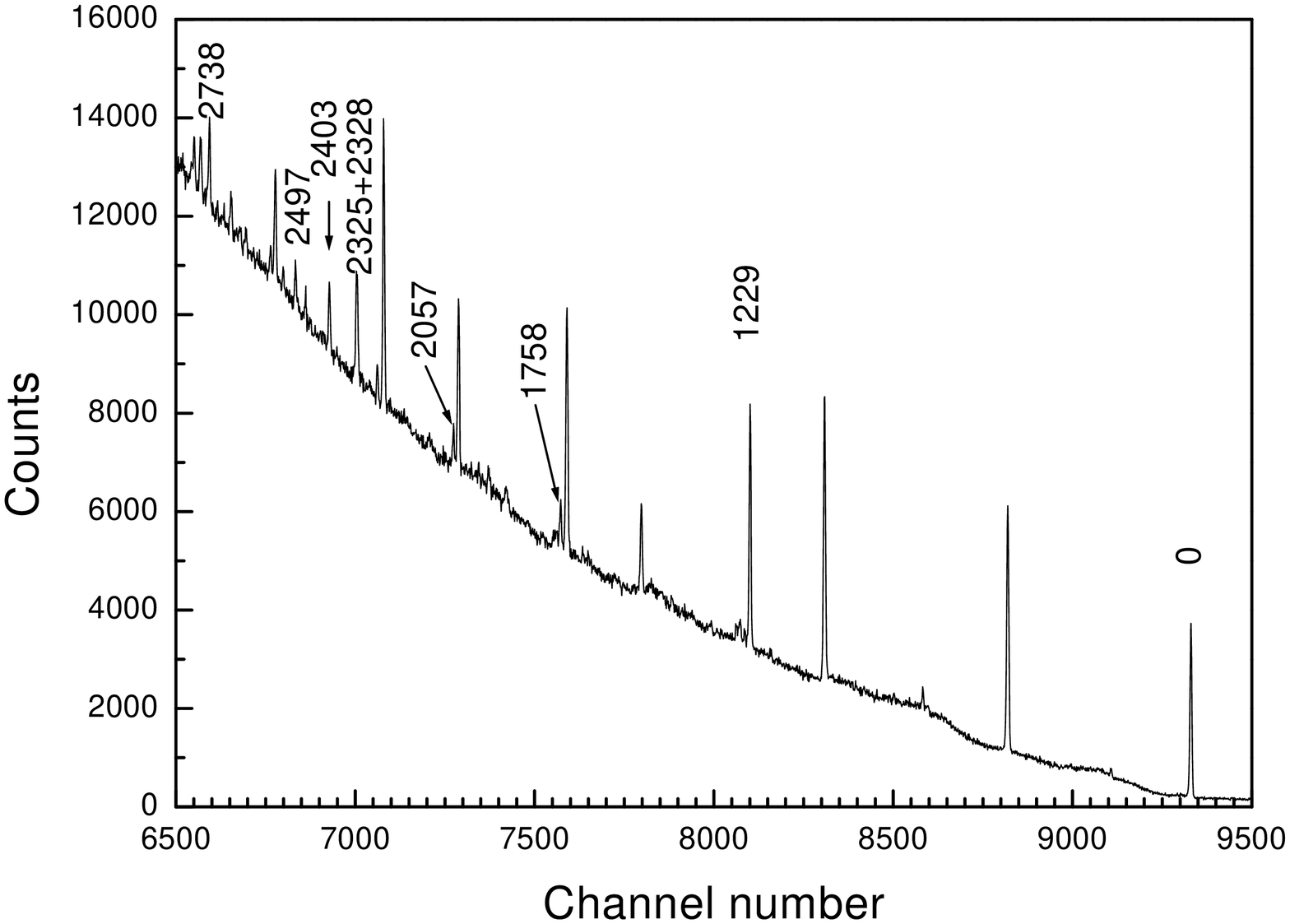}
\end{center}
\hspace{-0.8cm}

{\sl Fig.~1.~The part of the sum coincidence spectrum for the target
enriched in $^{117}Sn$. Full energy peaks are labelled with the energy (in keV)
of final cascade levels.}
\end{figure}

\newpage
\begin{figure}
\begin{center}
\leavevmode
\epsfxsize=11cm
\epsfbox{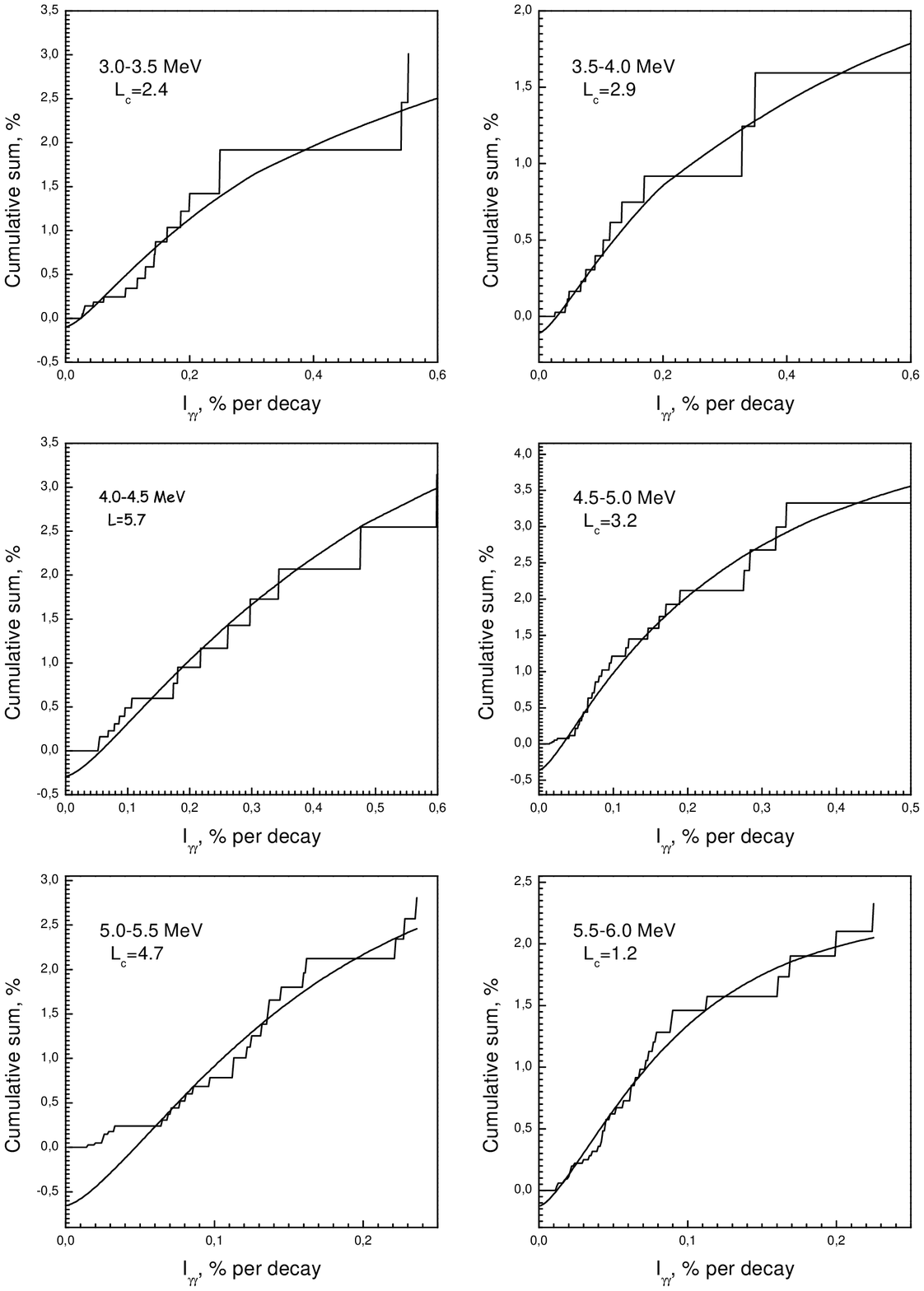}
\end{center}
\hspace{-0.8cm}

{\sl Fig.~2.~Lines represent approximation and extrapolation of the
cumulative intensities  of cascades $i_{\gamma\gamma}$ (Table~1)
in $^{118}Sn$ for 6 energy intervals of their intermediate levels
3.0-3.5, 4.0-4.5, 4.5-5.0, 5.0-5.5, 5.5-6.0 and 6.0-6.5 MeV
versus intensity. Histogram represents the experiment.
 $L_c$ is the approximated detection threshold (per $10^4$ decays).}
\end{figure} 

\newpage
\begin{figure}
\vspace{4cm}
\begin{center}
\leavevmode
\epsfxsize=11cm
\epsfbox{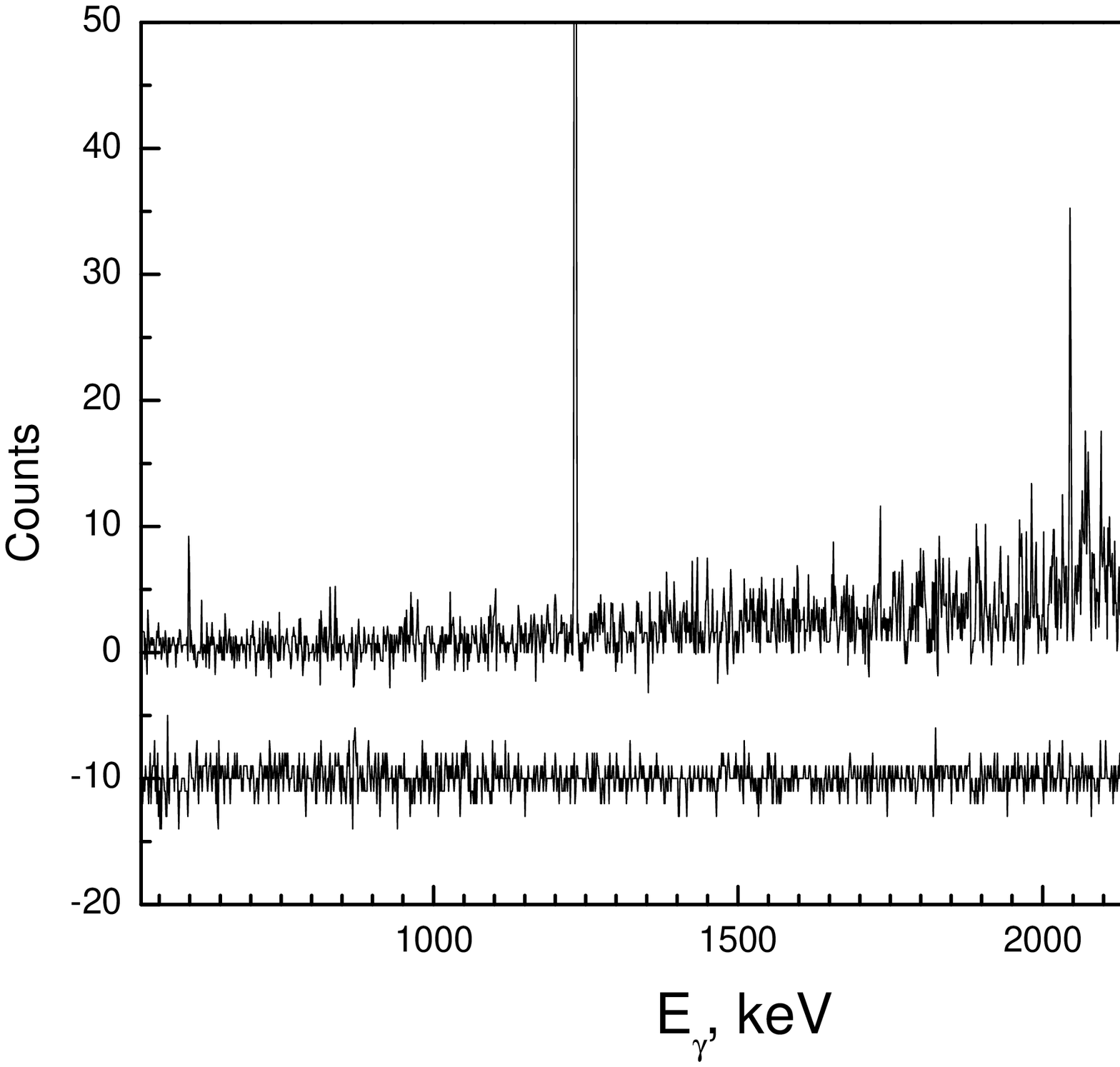}
\end{center}
\hspace{-0.8cm}
\vspace{-4cm}

{\sl Fig.~3.~The part of intensity distribution of the two-step cascades
terminating at the ground state of  $^{118}Sn$. The background spectrum with
the close sum energy presented for the comparison is shifted down by 10 counts.
}
\end{figure}
\begin{figure}
\begin{center}
\leavevmode
\epsfxsize=12.5cm
\epsfbox{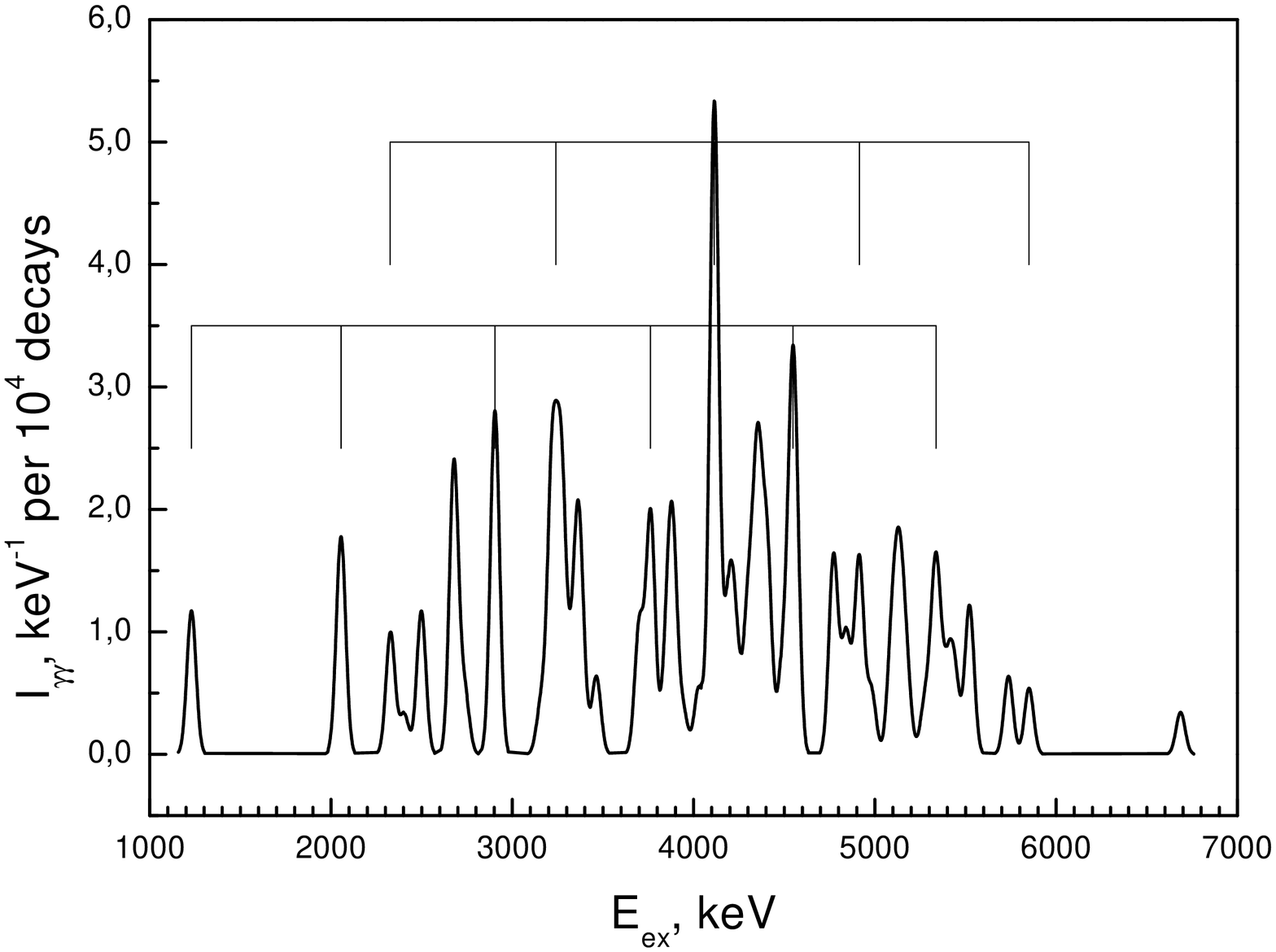}
\end{center}
\hspace{-0.8cm}

{\sl Fig.~4.~The dependence of the ``smoothed" intensities  of
resolved  cascades listed in Table 1 on the excitation energy. Possible
``bands" of practically harmonic excitations of the nucleus are marked.
The parameter $\sigma =25$ keV was used.}

\end{figure}

\begin{figure}
\begin{center}
\leavevmode
\epsfxsize=12.5cm
\epsfbox{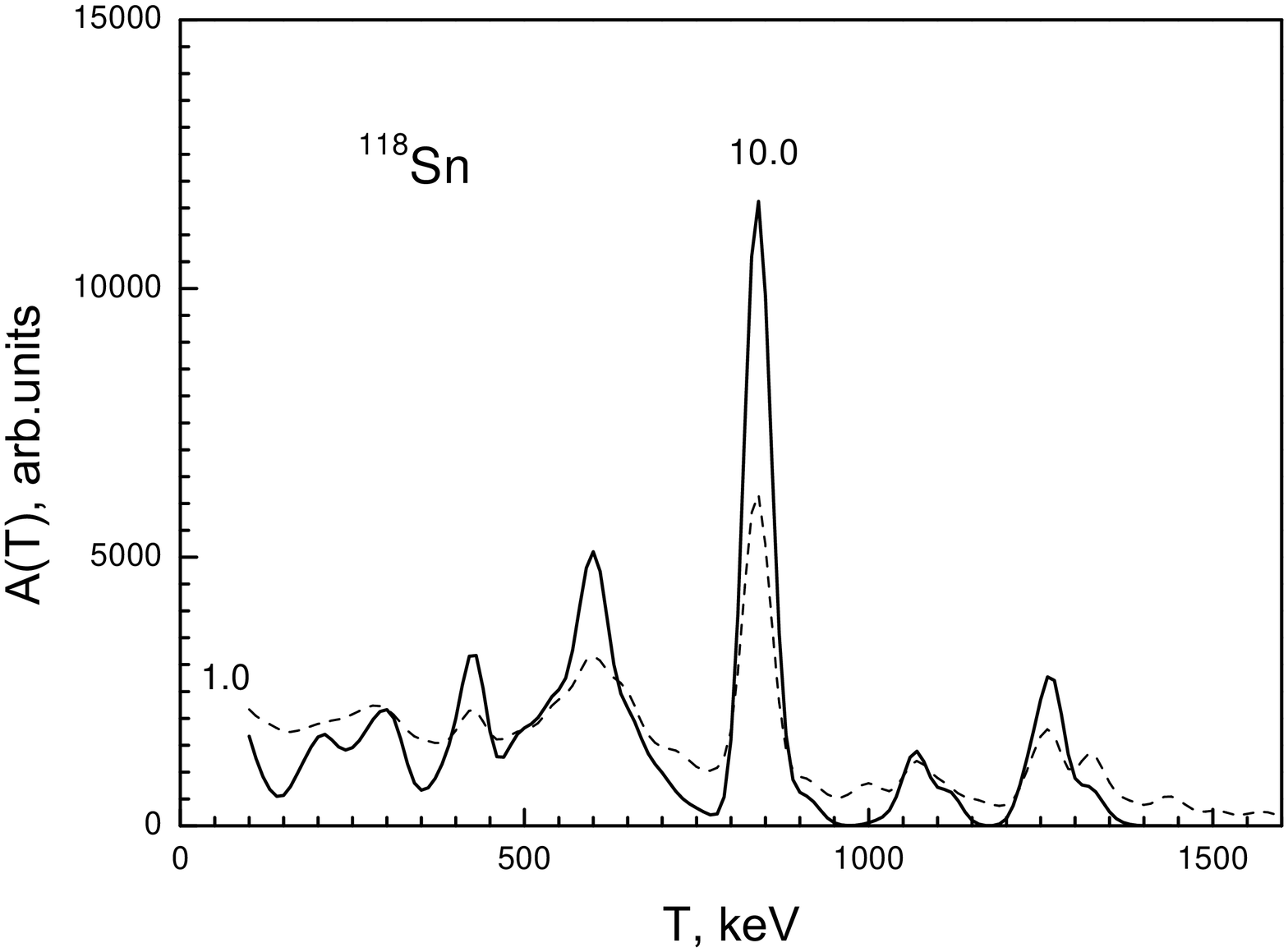}
\end{center}
\hspace{-0.8cm}

{\sl Fig.~5.~The values of the functional $A(T)$ for two registration
thresholds  of most intense cascades. The value of the registration
threshold (per $10^4$ decays) is given in the figure.}

\end{figure}

\begin{figure}
\begin{center}
\leavevmode
\epsfxsize=12.5cm
\epsfbox{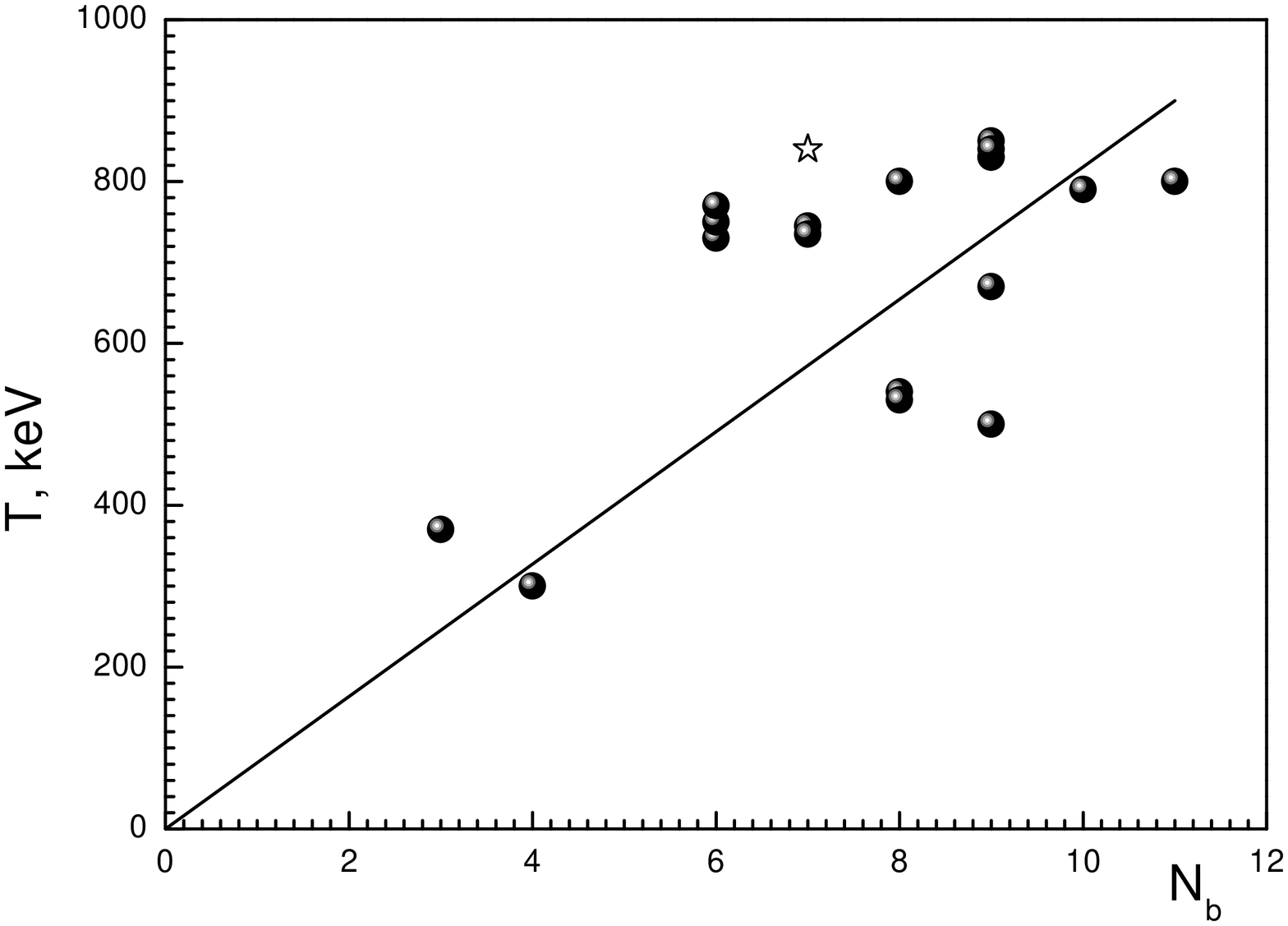}
\end{center}
\hspace{-0.8cm}

{\sl Fig.~6.~The value of the equidistant period $T$ for $^{118}Sn$
(asterisk) and other investigated even-even nuclei as a function of the
number of boson pairs $N_b$ in the unfilled shells. The line represents
possible dependence (drawn by eye).}
\end{figure}

\begin{figure}
\begin{center}
\leavevmode
\epsfxsize=12.5cm
\epsfbox{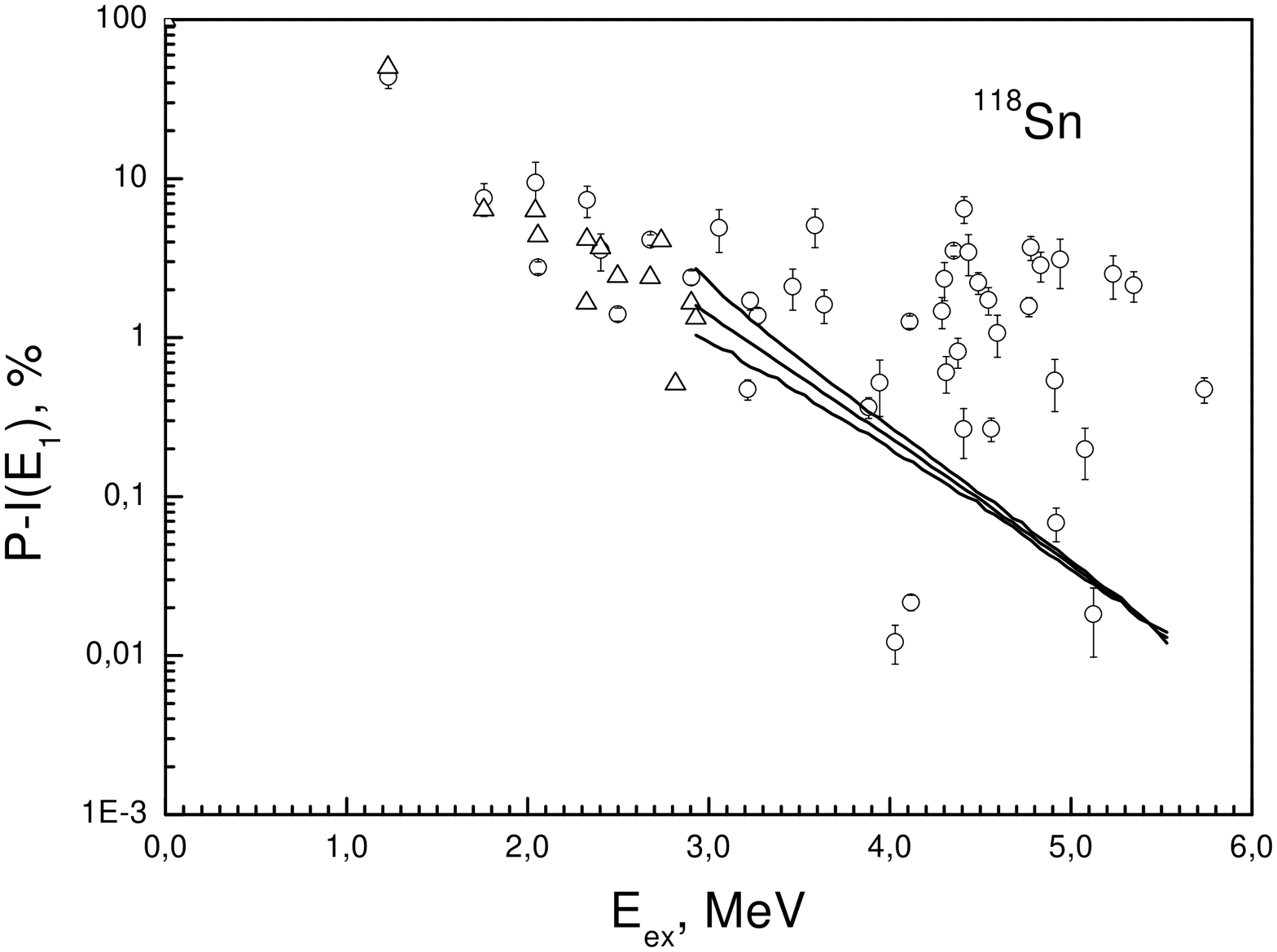}
\end{center}
\hspace{-0.8cm}
{\sl Fig. 7.~ Experimental population of different levels in $^{118}Sn$
(points with bars) in comparison with that calculated for levels 
$J^{\pi}=0^+, 1^+, 2^+$ within models [6] and [7] using $k(M1)=const$ - 
curves and triangles.}
\end{figure}
\end{document}